\documentclass{PoS}

\usepackage{calc}
\usepackage{graphicx}
\usepackage{amssymb,amsmath,amsthm,enumitem}
\usepackage{relsize}
\usepackage{multirow}
\usepackage{rotating}
\usepackage{bm}
\usepackage{enumitem}
\usepackage{url}
\usepackage{booktabs}
\usepackage{float}

\usepackage{setspace}
\usepackage{graphicx}
\usepackage{multicol}

\usepackage[normalem]{ulem}

\newcommand\be{\begin{equation}}
\newcommand\ee{\end{equation}}
\newcommand\bea{\begin{eqnarray}}
\newcommand\eea{\end{eqnarray}}

\title{
\begin{flushright}\begin{small}
 {\it  CP3-Origins-2018-038 DNRF90\\
  IFT-UAM/CSIC-18-99 \\
  FTUAM-18-21\\} \vspace{0.5cm}
  \end{small}
\end{flushright}
  On the definition of schemes for computing leading order isospin breaking corrections}

\ShortTitle{On the definition of schemes for computing  LO IB corrections}

\author{\speaker{A.~Bussone}$^{\,\,a,b}$, M.~Della Morte$^{\,c}$,
  T.~Janowski$\,^d$, A.~Walker-Loud$\,^{e,f}$\\
\llap{$^a$}{Department of Theoretical Physics, Universidad
Aut\'{o}noma de Madrid, E-28049 Madrid, Spain}\\
\llap{$^b$}{Instituto de F\'{i}sica Te\'{o}rica UAM-CSIC, c/
Nicol\'{a}s Cabrera 13-15, Universidad
Aut\'{o}noma de Madrid, E-28049 Madrid, Spain}\\
\llap{$^c$}{CP$^3$-Origins, University of Southern Denmark, Campusvej
55, 5230 Odense, Denmark}\\
\llap{$^d$}{School of Physics and Astronomy, University of Edinburgh,
Edinburgh EH9 3JZ, UK}\\
\llap{$^e$}{Nuclear Science Division, Lawrence Berkeley National
Laboratory, Berkeley, CA 94720, USA}\\
\llap{$^f$}{Physics Division, Lawrence Livermore National Laboratory,
Livermore, CA 94550, USA}\\
E-mail:\email{andrea.bussone@uam.es,dellamor@cp-origins.net,
t.janowski@ed.ac.uk,awalker-loud@lbl.gov}
}

\abstract{
  We discuss an alternate scheme, or a 'line of constant physics', which can be used 
  when computing
  isospin breaking corrections to hadronic quantities.
  We show that within a certain class of schemes
  one can separate the electromagnetic corrections from the
  strong isospin breaking corrections at
  leading order, meaning that within this class scheme-ambiguities are
  higher order effects.
}

\FullConference{The 36th Annual International Symposium on Lattice Field Theory - LATTICE2018\\
		22-28 July, 2018\\
		Michigan State University, East Lansing, Michigan, USA.}

\begin{document}

\section{Introduction}
Lattice QCD has reached in many instances such a level of precision that the
breaking of isospin symmetry has become the main source of uncertainty
(see, for example~\cite{Tantalo:2013maa} and~\cite{Aoki:2016frl}).
This breaking has two origins; one is due to the difference in the electric
charge of the light quarks (typically indicated as ``EM'' or ``QED'' breaking)
and the other is due to their difference in mass (typically indicated as
``strong ispsopin'' breaking).
Isospin breaking from these two sources is intimately coupled as the electromagnetic interactions renormalize the quark mass operators, which in turn serve as counter-terms to ultra-violet divergences arising from radiative QED corrections.
Therefore, assessing the amount isospin breaking arising from QED or strong sources, and the underlying Lagrangian parameters, is necessarily renormalization scheme dependent.

In order to obtain results with accuracies beyond the percent level, as needed
for example for the hadronic contribution to the anomalous magnetic moment of
the muon, lattice calculations have to depart from the isospin symmetric limit.
One possible approach consists in ``simply'' simulating QED+QCD with $N_f$
quarks of different masses. Such a theory has $N_f+2$ parameters, and assuming
it gives an accurate description of Nature (at low energies, perhaps), one can
fix those parameters by choosing $N_f+2$ ``reasonably independent''
experimental inputs. Upon choosing a renormalization scheme, these bare
parameters are then converted to renormalized ones.
Any other choice of $N_f+2$ ``reasonably independent'' experimental inputs would
produce the same renormalized parameters, again assuming that QED+QCD with $N_f$
flavors gives an accurate description of Nature (and barring accidental
degeneracies). In this approach there is no ``scheme dependence'' (meaning
dependence on the choice of quantities used to fix the $N_f+2$ paramaters)
in the results.

However, for different reasons, one is often interested in estimating the size
of isospin symmetry breaking corrections, by imagining a Taylor expansion
of observables around the isospin symmetric point. 
For example, the primordial amount of ${}^4$He produced in Big-Bang-Nucleosynthesis is very sensitive to isospin breaking~\cite{Heffernan:2017hwa}.
Quantitatively assessing the dependence from the two sources of isospin breaking allows for strong constraints to be placed upon
the possible time-variation of this fundamental constant.
We elaborate more
on the Taylor expansion mentioned above in this proceedings contribution, however, it is clear
that in comparing an ``isospin symmetric world'' to one which is not, one has
to specify what is kept fixed in the comparison and that produces a scheme
ambiguity on which we collect a number of remarks in the following.
We discuss in some detail, a class of schemes in which the scheme-ambiguities are second order in isospin breaking, allowing for a meaningful separation of these effects to a precision that is sufficient for current and projected lattice QCD applications.

\section{Setting up the expansion}
We consider a quantity $\mathcal{O}$ that we want to compute including QCD + IB (isospin breaking) corrections to leading order.
In order to setup an expansion (for the IB contributions), we need to write its dependence on ``independent'' parameters. One of the subtleties is related to
$\delta m = m_d-m_u$ as that is in principle a function of $\alpha$ (here
being the EM coupling).

We work in the simpler setup of $N_f=2$, where the main issues are however already present, and write $\mathcal{O}$ as a function of renormalized parameters (neglecting $O(\alpha)$ corrections to $\alpha_{strong}$)\footnote{
Those can be absorbed into a change in the lattice spacing. One can fix the relative lattice spacing by
computing $r_0/a$, as done in~\cite{QEDM_amu}, or any other ``gluonic'' quantity (e.g., $t_0$).
In the electroquenched approximation that is independent from $\alpha$ to all orders, as there is no direct 
coupling between photons and gluons. Beyond the electroquenched approximation one should extend the definition
of the scheme by keeping fixed a quantity which depends on $\alpha_s$ only up to quadratic corrections
in $\alpha$ and $m_d-m_u$.
}
\begin{equation}
\mathcal{O}=\mathcal{O}\left((m_d-m_u)_R(\alpha), (m_d+m_u)_R(\alpha), \alpha\right)\;,
\ee
with, say, $R=\overline{\rm MS}$ at 2 GeV. The arguments are clearly not independent, so, let us first of all
fix $(m_d+m_u)_R(\alpha)=(m_d+m_u)_{R,phys}=(m_d+m_u)_{PDG}\approx 7$ MeV~\cite{PDG2018}
for all values of $\alpha$. That makes
the parameter $\alpha$-independent by construction. In order to achieve that,
one can fix the sum of PCAC quark masses, which requires computing
$O(\alpha)$ corrections to renormalization factors as $Z_A$ and $Z_P$.

An alternative option is to  keep  $m_{\pi^0}^2$ fixed.
In $\chi$PT the EM corrections to $m_{\pi^0}^2$ are chirally suppressed and extremely small, at the 0.2\% level at NLO~\cite{Bij96,Bij2006},
or practically speaking, numerically second order in IB.
This follows from the observation that $\delta m \sim m_d+m_u$.
Strong IB breaking corrections to $m_{\pi^0}^2$ start at $O(m_d-m_u)^2$,
therefore keeping $(m_d+m_u)_R$ fixed is equivalent, at the order we are working, to keep $m_{\pi^0}^2$
fixed at its (physical) value as we change $\alpha$. 
Notice however that the neutral pion correlator outside the isospin symmetric limit receives quark-disconnected contributions (also chirally suppressed, see~\cite{Basak:2018yzz}),
the $\pi^0$ mixes with the $\eta$-like particles
and beyond the electroquenched approximation it decays into two photons
(although the width is very narrow)

Now one argument of $\mathcal{O}$ is fixed through one
of the options above, and to remind us we write
\be
\mathcal{O}=\mathcal{O}\left((m_d-m_u)_R(\alpha), 7 \;{\rm MeV}, \alpha\right)\;.
\ee
To proceed further we need the function $(m_d-m_u)_R(\alpha)$, i.e., the quark mass splitting
as a function of $\alpha$. That however is not uniquely defined, as we need to specify what we keep
constant (on top of $m_{\pi^0}^2$) to tune the masses as we change $\alpha$. We will look at two examples.

In the first let us say we keep fixed the neutron-proton splitting to its physical value.
Then, by looking at Fig.~3 of~\cite{abinitio}, qualitatively, one gets something like the red dashed line in
Fig.~\ref{fig:thefigure}.
\begin{figure}[thb]
\vspace{-0.3cm}
\begin{center}
\includegraphics[width=10.5cm]{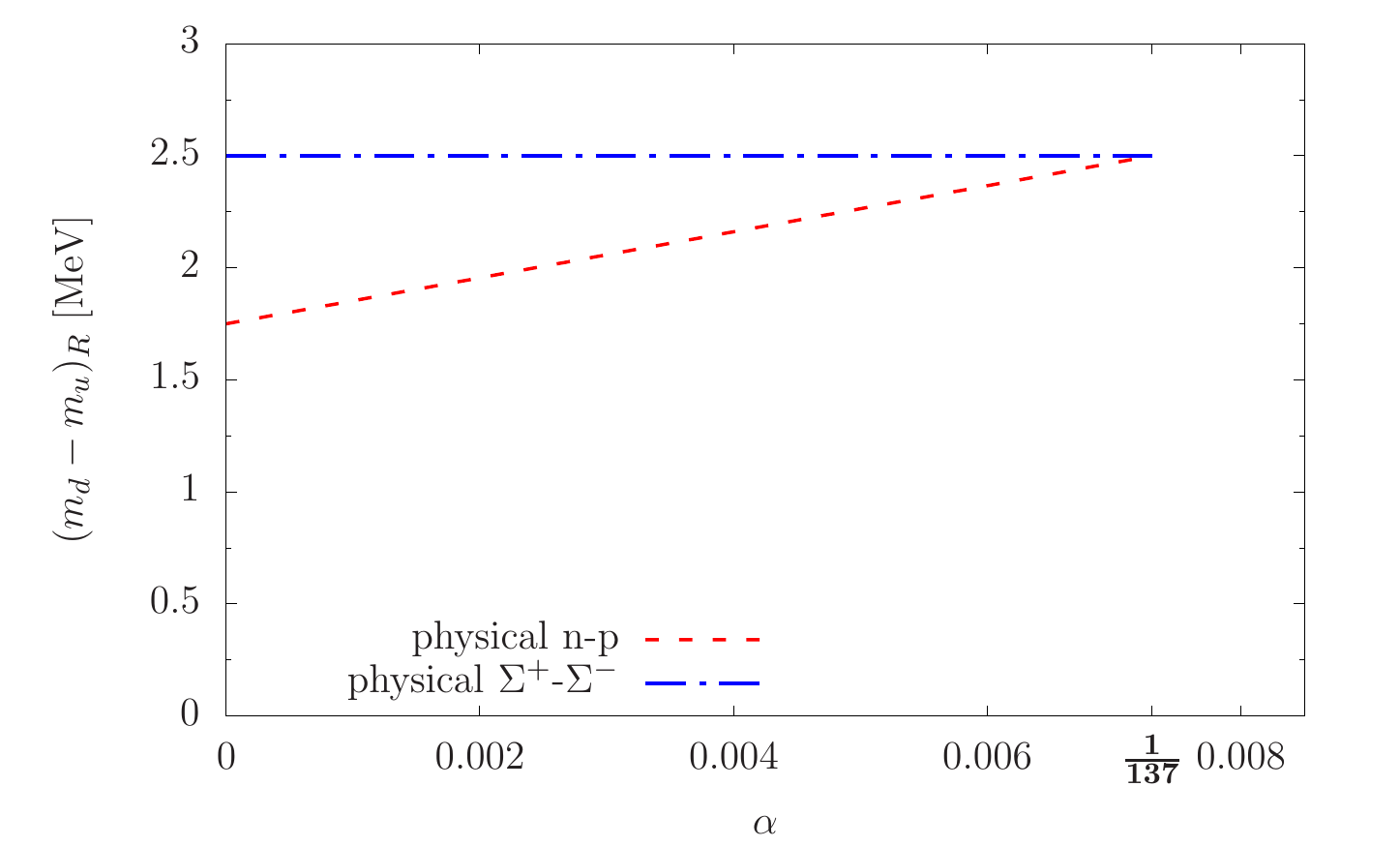}
\caption{Red dashed: qualitative behaviour of  $(m_d-m_u)_R(\alpha)$ as ``guessed'' from
 Fig. 3 in~\cite{abinitio} by requiring the neutron proton splitting to stay constant.
 2.5 MeV is roughly the physical (PDG or FLAG) value for the light quark mass difference. 
 Blue dot-dashed: Qualitative behaviour of  $(m_d-m_u)_R(\alpha)$ by requiring the 
$\Sigma^+-\Sigma^-$ splitting to stay constant.}
\label{fig:thefigure}
\end{center}
\vspace{-0.55cm}
\end{figure}
So a decrease of about 30\% at $\alpha=0$ with respect to the value at $\alpha=1/137$.

One can also use a different condition, perhaps yielding a smaller dependence of $(m_d-m_u)_R$ 
on $\alpha$. For example, as suggested in~\cite{abinitio}, one can keep the splitting between $\Sigma^+$ and $\Sigma^-$ fixed to
its physical value\footnote{We remind the reader, $\Sigma^+$ is a $uus$ baryon and $\Sigma^-$ a $dds$ one. We are assuming here that the (valence) strange quark mass has been fixed through an hadronic condition, for example by requiring
the combination $m_{K^0}^2+m_{K^+}^2$ to take its physical value.}. 
Since the two baryons have the same charge (in absolute value), the EM corrections to the masses are the same 
at leading order in $\alpha$ (neglecting structure-dependent contributions) 
and the splitting is due to the $m_d-m_u$ difference only (see also~\cite{Erben:2014hza} for a dispersive estimate).
One could therefore expect a dependence as depicted by the blue dot-dashed line in
Fig.~\ref{fig:thefigure}; a much weaker one (starting at $O(\alpha^2)$) compared to the previous case.

There is no inconsistency here. The only requirement for all choices is that at the physical
value of $\alpha$ all conditions give the same physical splitting\footnote{Strictly speaking that would happen
only for the six-flavor theory with physical masses, in the infinite volume ... }, but for $\alpha\neq 1/137$
one needs to specify what is meant by a world where $\alpha$ differs from its physical value,
since there are other parameters to be fixed (in this case the mass splitting).

So, some definitions of the function $(m_d-m_u)_R(\alpha)$ may be better than others in
some respect, but in prnciple they are all theoretically good.
Since we will be interested in $(m_d-m_u)_R(0)$, it makes sense to ask how much that changes
(parameterically) for two different definitions of the world with $\alpha\neq 1/137$.

Defining $\Delta_i m(\alpha)=\left.(m_d-m_u)_R(\alpha)\right|_{world \; i}$, and
dropping the index $R$ from now on (all masses should be understood as renormalized), the requirement we just discussed implies
\be
\Delta_1 m(\alpha=1/137) = \Delta_2 m(\alpha=1/137)= \Delta m_{phys}\;,
\label{eq:1}
\ee
and, by linearizing the dependence around $\alpha=1/137$
\be
\Delta_i m(\alpha)=\Delta m_{phys}+ \left(\alpha -\frac{1}{137} \right)*c_i \;.
\label{eq:2}
\ee
This linear hypothesis alone would give $\Delta_1 m(0)-\Delta_2 m(0)=O(\alpha_{phys})$,
which would translate into an ambiguity of $O(\alpha)$ once  inserted in the expansion of the quantity $\mathcal{O}$, as we shall see.
We need a stronger bound and for that we use the fact that in renormalized perturbation
theory the EM corrections to the mass of the quarks are multiplicative as a consequence of
residual chiral symmetries (axial and vector):
\be
m_{u,i}(\alpha)=m_{u,i}(0) Z_{u,i}(\alpha)\;, \quad {\rm and}\quad m_{d,i}(\alpha)=m_{d,i}(0) Z_{d,i}(\alpha)\;,
\label{eq:QEDren}
\ee
with $Z_{X,i}(\alpha)=1+C_{X,i}\alpha+\cdots$.
The relation above is among renormalized quantities; in particular the mass on the r.h.s. is
the renormalized mass in QCD and the coefficient $Z_X^i(\alpha)$ is expanded
in powers of the renormalized EM coupling. One can therefore imagine having
used a regularization preserving (part of) the chiral symmetry, where
indeed quark masses renormalize multiplicatively even at finite lattice spacing.
Notice also that the chosen ``scheme'' preserves the fact that the QED corrections to the quark
masses are multiplicative. In fact if one were to set the neutral pion mass to zero at $\alpha=0$,
which would enforce vanishing light quark masses, and then keep the same condition for all values of $\alpha$,
the light quarks would remain massless (at least to the order in the EM coupling that we are considering).\\ 
\indent The splitting now reads
\begin{eqnarray}
\Delta_i m (\alpha) &=& \Delta_i m(0)\, Z_{d,i}(\alpha)+ (Z_{d,i}(\alpha)-Z_{u,i}(\alpha))m_{u,i}(0) \;, \nonumber \\
&=&  \Delta_i m(0)\,(1+C_{d,i}\alpha) + C_{(d-u),i}\alpha m_{u,i}(0) \;.
\end{eqnarray}
For $\alpha=1/137=\alpha_{phys}$, through 
 eq.~\ref{eq:1}, we obtain
\begin{eqnarray}
&&\Delta_1m(0)(1+C_{d,1} \alpha_{phys})+C_{(d-u),1}\alpha_{phys}m_{u,1}(0)=  \nonumber \\
&&\Delta_2m(0)(1+C_{d,2} \alpha_{phys}) +  C_{(d-u),2}\alpha_{phys}m_{u,2}(0) \:,
\end{eqnarray}
which we rewrite as
\begin{eqnarray}
\Delta_1m(0)-\Delta_2m(0) &=& \alpha_{phys}\left( C_{d,2} \Delta_2m(0) - C_{d,1}\Delta_1m(0)\right) +   \; , \nonumber \\
&& \alpha_{phys}\left(  C_{(d-u),2} m_{u,2}(0)-   C_{(d-u),1} m_{u,1}(0)\right) \;.
\end{eqnarray}
That shows that the difference in $\Delta m$ in the two worlds is O($\alpha$) at least (we knew that already).
Similarly, by requiring the up-quark masses to be the same at $\alpha=1/137$, starting from eq.~\ref{eq:QEDren},
one concludes that the difference in the $m_{u,i}(0)$ is also $O(\alpha)$ at least.
The previous equation can then be cast in the form
\be
\Delta_1m(0)-\Delta_2m(0)=  O(\alpha^2) + O(\alpha \delta m) + O(\alpha m_u(0))\;.
\ee
Now,  {\it{using the fact that, numerically, $\delta m \simeq m_u$}} ($m_d$ and $m_u$ differ roughly by a factor 2), one obtains 
\be
\Delta_1m(0)-\Delta_2m(0)=  O(\alpha^2) + O(\alpha \delta m) \;,
\ee
which means that two different definitions of the $\alpha$-dependence of the up-down quark mass splitting
provide values for $\Delta m(0)$ that differ by higher orders in the IB-corrections. 

Coming back to the expansion of the observable $\mathcal{O}$, the conclusion is that one can either use
$\Delta_1 m$ or $\Delta_2 m$, as long as one is interested in the leading order corrections in
both $\alpha$ and $\Delta m$. We can now indeed rewrite $\mathcal{O}$ in terms of completely
independent variables by using $\Delta_1 m$ as first argument:
\be
\mathcal{O}=\mathcal{O}(\Delta_1 m(0), 7\,{\rm{MeV \,(or\,}}m_{\pi^0}^2\,{\rm fixed\, to \, its\, physical \, value)},\alpha)\;, 
\ee
which can be expanded as
\begin{eqnarray}
\mathcal{O} &=& \mathcal{O}(0, 7 \, {\rm MeV}, 0)+ \alpha_{phys} \left.\frac{\partial \mathcal{O}(0,7 \,{\rm MeV},\alpha)}{\partial \alpha}\right|_{\alpha=0} + \nonumber \\
  &+& \Delta_1 m(0) \left. \frac{ \partial\mathcal{O}(\Delta m,7 \,{\rm MeV},0)}{\partial \Delta m} \right|_{\Delta m=0} +  O(\alpha^2) + O(\alpha \delta m) \;.
\label{eq:expansion}
\end{eqnarray}
The first term should be computed in pure QCD with degenerate up and down quarks (with masses summing up to 7 MeV or $m_{\pi^0}=m_{\pi^0,phys}$).
For the second one needs to simulate QCD+QED tuning the bare masses for the up and down quarks, such that at each value of $\alpha$ 
the renormalized masses are the same.
Again the sum of the quark masses must be kept fixed for all values of $\alpha$.
There is an ambiguity here in the way the renormalized masses of the up and down quark are fixed to be degenerate
at $\alpha \neq 0$. If the renormalization factors are known including $O(\alpha)$ corrections, one can tune the bare
masses as described above. Otherwise one can adjust them such that the $\Sigma^+$ and the $\Sigma^-$ are degenerate,
or, following~\cite{Basak:2018yzz}, one can require the (squared) effective masses extracted from the connected diagrams
in the two-point functions  of the pseudoscalar density made either of up-quark fields only or of down-quark field only to be the same.
In both cases one is requiring a quantity proportional to the mass splitting up to corrections quadratic in the IB parameters to vanish\footnote{For
  the condition used in~\cite{Basak:2018yzz}, the corrections are actually $O(\alpha)$, but they are chirally suppressed.}.
It is easy to see then, that the ambiguity on the second term on the r.h.s of eq.~\ref{eq:expansion} is also quadratic in the IB parameters.

Finally, the third term can be computed in QCD with non-degenerate up and down quarks (and the by now usual
requirement on the sum of the masses). The ambiguity here comes from the definition of $\Delta m$, but
as we have argued in this contribution such ambiguity is second order in IB-corrections.

In conclusion, assuming all derivatives to be of $O(1)$, we see that all the ambiguities in the leading order expansion are
of higher order and in particular
we see that using $\Delta_2 m(0)$ instead of $\Delta_1 m(0)$
would change the result by $O(\alpha^2)$ and $O(\alpha \delta m)$, which we are anyhow neglecting.
Let us also remark that since $\Delta_1 m(0)$ differs by about 30\% from the physical splitting
(as one can infer from~\cite{abinitio}), as long as we are interested in IB corrections with
an accuracy of that order (which is reasonable), we could use the physical, PDG (or FLAG), $m_d-m_u$
value in eq.~\ref{eq:expansion}  instead of some $\Delta_im(0)$.
Notice also that all quantities in that equation are renormalized and the constraints are defined through renormalized quantities,
so in principle each term can be independetly computed and extrapolated to the continuum and infinite volume limit.

\section{Summary and conclusions}
We have shown that in separating strong from EM isospin breaking effects at leading order, a rather natural class of
schemes exists such that all ambiguities are of next-to-leading order. What charcterizes the schemes are the following
conditions:
\begin{itemize}
\item[{\it i)}] Throughout the different computations, one keeps fixed a quantity that depends on $m_u+m_d$ up
  to quadratic IB corrections (e.g. $m_{\pi^0}^2$), and in principle a quantity which at the same order only depends
  on $\alpha_{strong}$ (for setting the scale). For the latter, in the electroquenched approximation one can use $r_0$ extrapolated
  to the chiral limit.
\item[{\it ii)}] The condition fixing the mass splitting for $\alpha \neq 0$ should be smooth in $\alpha$. In particular we have used
  ``$\alpha$-independent'' conditions. The requirement that all such conditions provide the same splitting for $\alpha=1/137$,
  follows
  from the renormalizability of the theory, as discussed in the Introduction.
\item[{\it iii)}] The mass-symmetric point at $\alpha \neq 0$ is defined through a quantity which
  is proportional to the mass splitting and has no $O(\alpha)$ corrections, rather $O(\alpha \delta m)$ or $O(\alpha m_\ell)$.
\end{itemize}
In discussing these schemes and their ambiguities, we have often exchanged
$O(m_u)$ or $O(m_d)$ with $O(\delta m)$, e.g, when neglecting the chirally suppressed
$O(\alpha)$ corrections to $m_{\pi^0}^2$ or in point $iii)$ above. While this seems reasonable close to the physical point,
it is not clear to us how reliable that procedure is when working with pions with masses between 200 and 400 MeV,
as it is often the case in lattice computations. Exploring such effects is left for future studies.

\section*{Acknowledgments}
We acknowledge insightful conversations with C.~Sachrajda concerning ``hadronic'' schemes for determining QED and strong isospin breaking corrections.  See also the presentation at Lattice 2018~\cite{CS2018}.
A.B.~acknowledges support through the
Spanish MINECO project
FPA2015-68541-P,
the Centro de Excelencia Severo Ochoa Programme
SEV-2016-0597
and the Ram\'{o}n y Cajal Programme RYC-2012-10819.

\end{document}